\documentclass{article}

\usepackage{arxiv}
\usepackage{amsmath}
\usepackage{caption}
\usepackage{tabularray}
\usepackage[utf8]{inputenc} 
\usepackage[T1]{fontenc}    
\usepackage{hyperref}       
\usepackage{url}            
\usepackage{booktabs}       
\usepackage{amsfonts}       
\usepackage{nicefrac}       
\usepackage{microtype}      
\usepackage{lipsum}
\usepackage{graphicx}
\graphicspath{ {./images/} }

\title{Discovering Multi-omic Biomarkers for Prostate Cancer Severity Using Machine Learning}

\author{
 Jefferson Zhou \\
  Rye Country Day School\\
  \texttt{jefferson.zhou@gmail.com} \\
  \AND
  Kahn Rhrissorrakrai \\
  IBM Research, Yorktown Heights, NY USA \\
  \texttt{krhriss@us.ibm.com} \\
}

\begin{document}
\maketitle

\begin{abstract}
    
Prostate cancer is the second most common form of cancer, though most patients have a positive prognosis with many experiencing long-term survival with current treatment options. Yet, each treatment carries varying levels of intensity and side effects, therefore determining the severity of prostate cancer is an important criteria in selecting the most appropriate treatment. The Gleason score is the most common grading system used to judge the severity of prostate cancer, but much of the grading process can be affected by human error or subjectivity. Finding biomarkers for prostate cancer Gleason scores in a quantitative, machine-driven approach could enable pathologists to validate their assessment of a patient cancer sample by examining such biomarkers. In our study, we identified biomarkers from multi-omics data using machine learning, statistical tools, and deep learning to train models against the Gleason score and capture the most important features that could potentially serve as biomarkers for the Gleason score. Through this process, multiple genes, such as \textit{COL1A1} and \textit{SFRP4}, and cell cycle pathways, such as G2M checkpoint, \textit{E2F} targets, and the \textit{PLK1} pathways, were found to be important predictive features for particular Gleason scores. The combination of these analytical methods shows potential for more accurate grading of prostate cancer, and greater understanding of biological processes behind prostate cancer severity that could provide additional therapeutic targets. 
\end{abstract}


\section{Introduction}
Prostate cancer is the second most common form of cancer, where 6 in 10 prostate cancer patients are above the age of 65 \cite{noauthor_key_nodate}. Standard care treatments include surgeries, e.g. prostatectomy, and therapies targeted at cancer cells such as radiation and cryotherapy \cite{noauthor_cryotherapy_nodate,noauthor_radiation_nodate, noauthor_surgery_nodate}. Although these treatments are effective, given the long term survival of most prostate cancer patients, they may harm the patients' quality of life and can be unnecessarily severe in those cases where only more mild treatments are required \cite{noauthor_key_nodate}. To provide more fitting treatments commensurate with disease severity, prostate cancer needs to be better understood and modeled more accurately.

The severity of prostate cancer is measured using the Gleason score. Gleason scores are determined by a pathologist assessing a tissue sample and assigning a primary and secondary grade from 1 to 5 based on how aggressive the cells appear, with 1 being the least severe and 5 being the most \cite{noauthor_prostate_2012, noauthor_gleason_nodate}. Primary and secondary grade patterns of less than 3 are rare, thus the addition of the two grades, which forms the final Gleason score, generally falls between 6 and 10. Although the Gleason score is a valuable clinical tool, the process of measuring the Gleason score can be made more accurate. Studies have shown that many structures in prostate cancer can alter the Gleason score, leading to over- or undergrading \cite{mckenney_histologic_2016, noauthor_study_2017}. 
Variability in the prediction of the Gleason score shows that this process is subject to human errors. To reduce such errors, machine learning can be used to predict the Gleason score in a more accurate and reliable manner~\cite{kartasalo_artificial_2021}. An explainable machine learning approach will also be able to identify important features and discover biomarkers that may explain prostate cancer severity or grading.

Machine learning and deep learning methods have been used to predict the Gleason score with reasonable success, though these efforts have been focused primarily on image analysis. One such study used radiomic features coupled with a Random Forest classifier~\cite{chaddad_predicting_2018}. Specifically, MRI imaging was used to find regions of interest, and radiomic features were extracted from the regions to predict the Gleason score using the random forest classifier.
The results of the study showed high accuracy (57.89\% - 84.00\%) across all folds and noted significant importance in two radiomic features, entropy and sum entropy. These results were consistent with previous studies where the entropy correlated with the Gleason score, and was also consistent with other studies that viewed the Gleason score as the default indicator between benign and malignant prostate cancer. 

Deep learning technology has also been used to predict the Gleason score. In one study, a two-stage deep learning system was developed \cite{nagpal_development_2019}. In the first stage, the model was trained to predict the Gleason pattern, and in the second stage, the pattern prediction was used to predict the Gleason grade group (1,2,3,4,5). With a validation dataset of 331 images from their patient cohort, they found that this deep learning approach 
had a higher accuracy of 0.70 compared to the mean accuracy of 0.61 for pathologists independently grading the same images and also had a lower mean average error when predicting Gleason patterns. 
Other studies using imaging of prostate cancer have also found deep learning approaches are effective at predicting the Gleason score and can potentially be assistive tools for both analyzing biopsies and improving prostate cancer diagnosis, especially when higher level expertise is not available~\cite{nagpal_development_2020,singhal_deep_2022}.

These prior works focused on a single data modality. In our study, we consider multiple data modalities, specifically multiple omic data types, to improve Gleason score prediction and to identify biomarkers. Our contributions are in two aspects. First, we leveraged whole-exome sequencing and RNA-seq data from The Cancer Genome Atlas (TCGA) Program. Each of these omics datasets can have information that is domain specific, so models using multiple omics datasets together can potentially find biomarkers that maximize information from across modalities. Second, we focused on two different machine learning techniques. The first is a Random Forest (RF) model to identify features that are important for predictions. These random forest models have shown themselves highly effective when analyzing multi-modal biological data. The second is a deep learning method for modeling gene expression data, Transformer for Gene Expression Modeling (T-GEM), that was developed for predicting cancer types \cite{zhang_transformer_2022}. By using machine learning models to predict the Gleason score, important biomarkers, whether as single gene markers, gene sets, or signatures, can be identified to provide potentially more consistent prostate cancer grading as well as additional therapeutic targets.

\section{Methods and Procedures}

\subsection{Data pre-processing}
Prostate cancer data from The Cancer Genome Atlas (TCGA) was downloaded from the Genomic Data Commons (GDC) on April 4th, 2022\cite{abeshouse_molecular_2015}\cite{noauthor_gdc_nodate}. We downloaded pre-processed RNA-Seq (transcripts per million (TPMs)), gene-level copy number (CN), and mutation annotations (MuTect2 VCFs) data, as well as relevant clinical and sample information.
RNA-seq data was $log_2$ transformed with a $+1$ pseudocount. CN data was $log_2$ normalized, $cn_{log} = log_{2}(cn/2)$, where $cn$ is the measured copy number. Mutations were filtered to exclude silent mutations and mutations in the intron, 5'UTR, or 3'UTR. The mutation data were processed at the gene level as the absolute mutational load per gene after filtration. Patient samples were filtered for those with complete clinical, genomic, and transcriptomic data. The feature space for each of the modalities varied: RNA-seq data - 19938 genes, mutation data  - 18701 genes, and CN data - 59104 elements, including entities such as genes, pseudogenes, miRNAs, etc.
Two methods were used to filter the gene feature space. The first method used common cancer genes from the Cancer Gene Census to subselect the gene space\cite{sondka_cosmic_2018}. The second method used a z-score of the feature importance value from the RF model that was greater than a given threshold. Filtration method was dependent on the experiments performed. Gene sets for the Hallmark and C2CP gene set collections were downloaded from the GSEA Human Molecular Signatures Database website in July 2023 \cite{noauthor_gsea_nodate}.

\subsection{Random Forest classifier}
The random forest algorithm was used as a classifier to predict the Gleason score per sample using gene-level features as input. The \textit{sklearn} package (version 1.0.2) \textit{random\_forest\_classifier} was used with the default parameters (n\_estimators=100, criterion="gini", max\_depth=None, min\_samples\_split=2, min\_samples\_leaf=1) along with a random seed of 6. Binary classification setting and five-fold cross validation was used for model training and testing. 
For cross validation, the k-fold shuffle parameter was set to true and the random seed was set to 6. To find the importance values of each gene as assigned by the RF, the Gini importance value from the \textit{feature\_importances\_} attribute of the model was used~\cite{noauthor_sklearn.ensemble.randomforestclassifier_nodate}. Genes and their importance values were retained for each of the five folds of the experiments and then averaged to order genes from highest to lowest importance. Average importance values were filtered using a z-score threshold for better interpretability during analysis.

\subsection{Gene set analysis}

The \textit{prerank} function from the \textit{gseapy} module (version 1.0.4) was used to identify gene pathway significance for each Gleason score in the dataset. Parameters: \textit{rnk} = mean values of each feature for specified Gleason score in dataset sorted in descending order, \textit{gene sets} = hallmark pathway database or C2CP pathway database, \textit{minimum gene set size} = 10, \textit{maximum gene set size} = 500, \textit{permutation number} = 1000, \textit{seed} = 6. The full dataset was used without filtering when performing GSEA.
GSEA was performed in two experimental designs: pairwise (e.g. 6 verses 9) or one-vs-all (e.g. 6 vs 7, 8, 9). Positive NES scores would mean up-regulation in the higher Gleason score for pairwise comparisons and up-regulation in the isolated Gleason score for one-vs-all comparisons.

The \textit{scipy.stats} (version 1.7.3) \textit{hypergeom.sf} function was used to calculate significance for gene sets in relation to important genes found by the random forest classifier in various single-omic experiments. Parameters include the intersection between the gene set and important gene list (\textit{k}), intersection between gene space and the chosen gene database (\textit{N}), the size of the gene set (\textit{n}), and the size of the important gene list (\textit{M}).

\subsection{T-GEM analysis}
The T-GEM (Transformer for Gene Expression Modeling) model is a novel, interpretable deep learning model primarily focused on gene expression data~\cite{zhang_transformer_2022}. The utilization of self-attention that is characteristic of transformer models enables T-GEM to model unordered input like gene expression data and learn gene-gene interactions. Furthermore, the property of self-attention to make a new representation of a word within the entire context of a word sequence is able to be applied to finding the importance of a gene in the context of prostate cancer severity.
The T-GEM code was downloaded from https://github.com/TingheZhang/T-GEM on August 9th, 2023 ~\cite{zhang_transformer_2022}. 
Settings included a \textit{head size} of 5, \textit{batch size} of 1, \textit{dropout rate} of 0.3, \textit{learning rate} of 0.0001, and an \textit{epoch size} of 30. Input genes were filtered for the top 2000 by variance.

\section{Results}

\subsection{Characterizing the TCGA Prostate Adenocarcinoma Dataset}
The Cancer Genome Atlas (TCGA) project provides a comprehensive dataset of prostate cancer with multiple data types and corresponding clinical information, such as the Gleason score, for 500 patients\cite{abeshouse_molecular_2015}.

\begin{figure}[!t]
\centering
\includegraphics[width=3in]{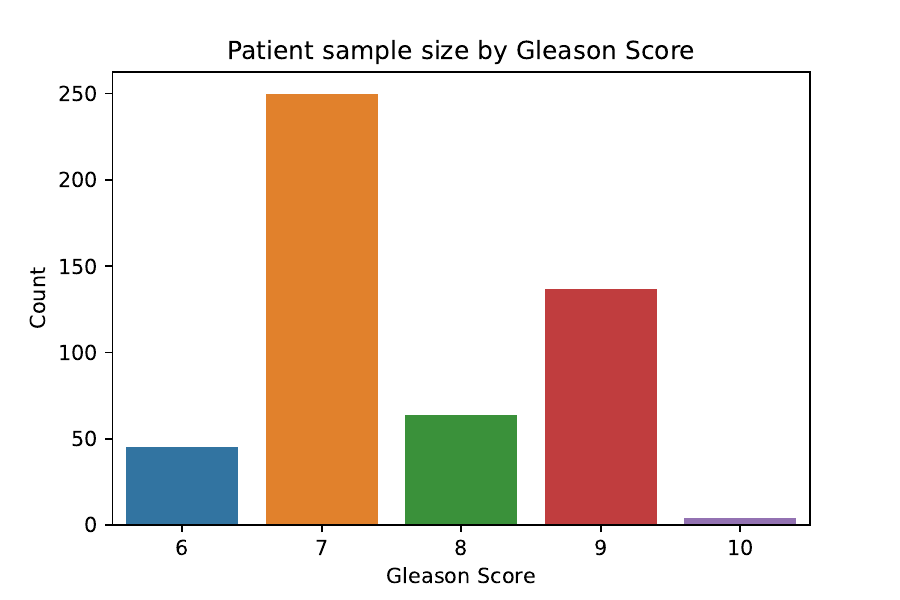}
\caption{ Distribution of TCGA prostate cancer patient samples by Gleason score.}
\label{fig:gleasonHist}
\end{figure}

We observed that the patient Gleason score distribution was quite imbalanced (Figure \ref{fig:gleasonHist}) with Gleason 7 representing half of the cohort and Gleason 9 a third. There were too few patients with a Gleason score of 10, and these patients were excluded from subsequent analyses.
The number of elements and type of information each omics dataset contains vary, which may affect model performance.
We performed experiments using all features for a given data type, as well as sub-selecting for 576 common cancer genes from the Cancer Gene Census (CGC) to reduce the gene space size \cite{cosmic_cancer_nodate}. By reducing the gene feature space, we aimed to reduce the complexity and increase interpretability of the predictive models.

\subsection{Random forest performance with single data modalities}

\begin{figure}[!t]
\centering
\includegraphics[width=175mm]{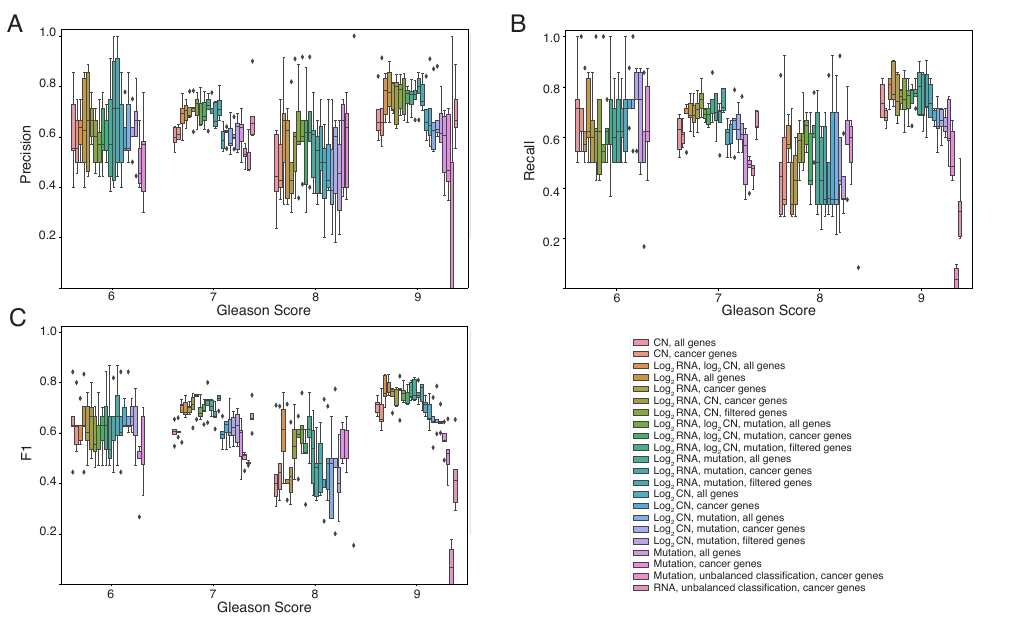}
\caption{ Summary performance across classification experiments. For each experiment in the legend, the input data, including whether RNA TPM values or copy number (CN) values were $log_2$ transformed, and gene feature space is indicated.  Unless otherwise specified all results were from experiments accounting for class imbalance. Each boxplot is the distribution across 5-folds for precision (A), recall (B), and F1-score (C).}
\label{fig:SummaryPerf}
\end{figure}

\begin{figure}[!t]
\centering
\includegraphics[width=165mm]{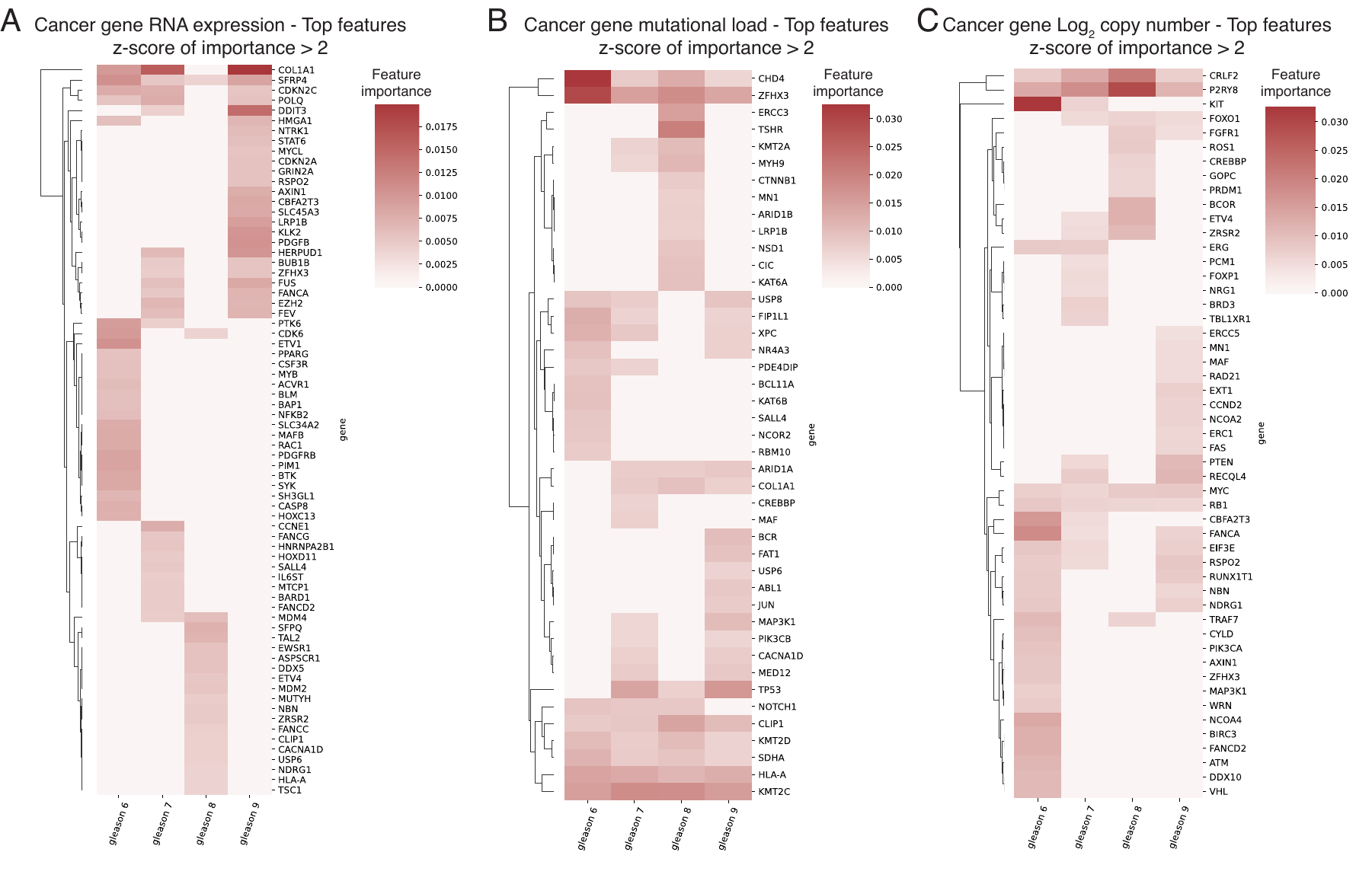}
\caption{Top features from a balanced binary classification using random forest. Top features were identified as those with a feature importance $z$-score $\ge 2$. Cells are colored according to their feature importance score. Top features from RF analysis of A) cancer gene $log_2$ RNA expression,  B) cancer gene mutational load, and C) cancer gene $log_2$ copy number.}
\label{fig:imp_heatmaps}
\end{figure}

We first evaluated the performance of the random forest model when only using one data type. Mutation data filtered with cancer genes gave the highest median F1-score of 0.67 when predicting Gleason 6, though copy number data and RNA data performed similarly (Figure \ref{fig:SummaryPerf}). For Gleason 7, RNA data had the highest performance, achieving a median F1-score of 0.70 when filtered with cancer genes. Copy number data exhibited similar performance, and mutation data had the worst performance. For Gleason 8, no single-omic modality performed well. Lastly, for Gleason 9, RNA data achieved the highest performance of any random forest experiment, with a median F1-score of 0.8 without any filters (Figure \ref{fig:SummaryPerf}). Overall, mutation data had the highest performance for Gleason 6, RNA data had the highest performance for Gleason 7 and 9, and copy number data was generally comparable to the top performer for each Gleason score.
Filtering for cancer genes slightly improved performance depending on the data type and Gleason score being predicted. 
For individual Gleason scores between 6 and 9, \textit{CHD4}, \textit{ZFHX3}, \textit{KMT2C}, \textit{TSHR}, and \textit{TP53} were cancer genes that had some of the highest feature importance (Figure \ref{fig:imp_heatmaps}). Using RNA data filtered by cancer genes, \textit{SFRP4}, \textit{COL1A1}, \textit{DDIT3}, \textit{ZFHX3}, \textit{CBFA2T3}, and \textit{POLQ} had the highest feature importance for predicting Gleason scores. Lastly, \textit{KIT}, \textit{CBFA2T3}, and \textit{FANCA} were found as the top features from copy number experiments when predicting Gleason 6. However, there were no significantly important genes for predicting Gleason 7 or 9 when using copy number data filtered by cancer genes.

\subsection{Performance of multi-omics models}

We then tested whether combining multiple data modalities would improve predictive performance. Models given “filtered genes” only utilized genes from previous single data modality cancer gene filtered experiments that had a feature importance value z-score $\ge 2$.  
We found, using mutation data and CN data with z-score filtered cancer genes, the RF achieved a median F1-score of 0.71 for predicting Gleason 6 (Figure \ref{fig:SummaryPerf}). Using RNA and CN data with filtered cancer genes, the RF reached an F1 of 0.75 for Gleason 7, though combining RNA and mutation data with filtered cancer genes achieved similar levels of performance. A majority of the multi-omics models performed poorly when predicting Gleason 8. The highest median F1-score was 0.62, where the features were either RNA and CN data with all genes, or a combination of all three omics datasets with either all genes or filtered cancer genes. Most multi-omics models had high performance when predicting Gleason 9 with the exception of those using mutation and CN data. Using RNA and CN data with either all cancer genes or z-score filtered cancer genes gave the best performance for Gleason 9.

Overall, models predicting Gleason 9 achieved the highest performance when combining data types, with models predicting Gleason 7 having the second highest performance. While the highest multi-modal model performance under-performed the highest performing single-modal model, overall model performance was higher for all Gleason scores, particularly when predicting Gleason 8. However, the random forest's ability to predict Gleason 8 remained lower as compared to its predictive performance for the other Gleason scores. We found that z-score filtered cancer genes gave better performance for predicting Gleason 7, but overall the use of cancer genes did not significantly affect performance.

\begin{figure}[!t]
\centering
\includegraphics[width=145mm]{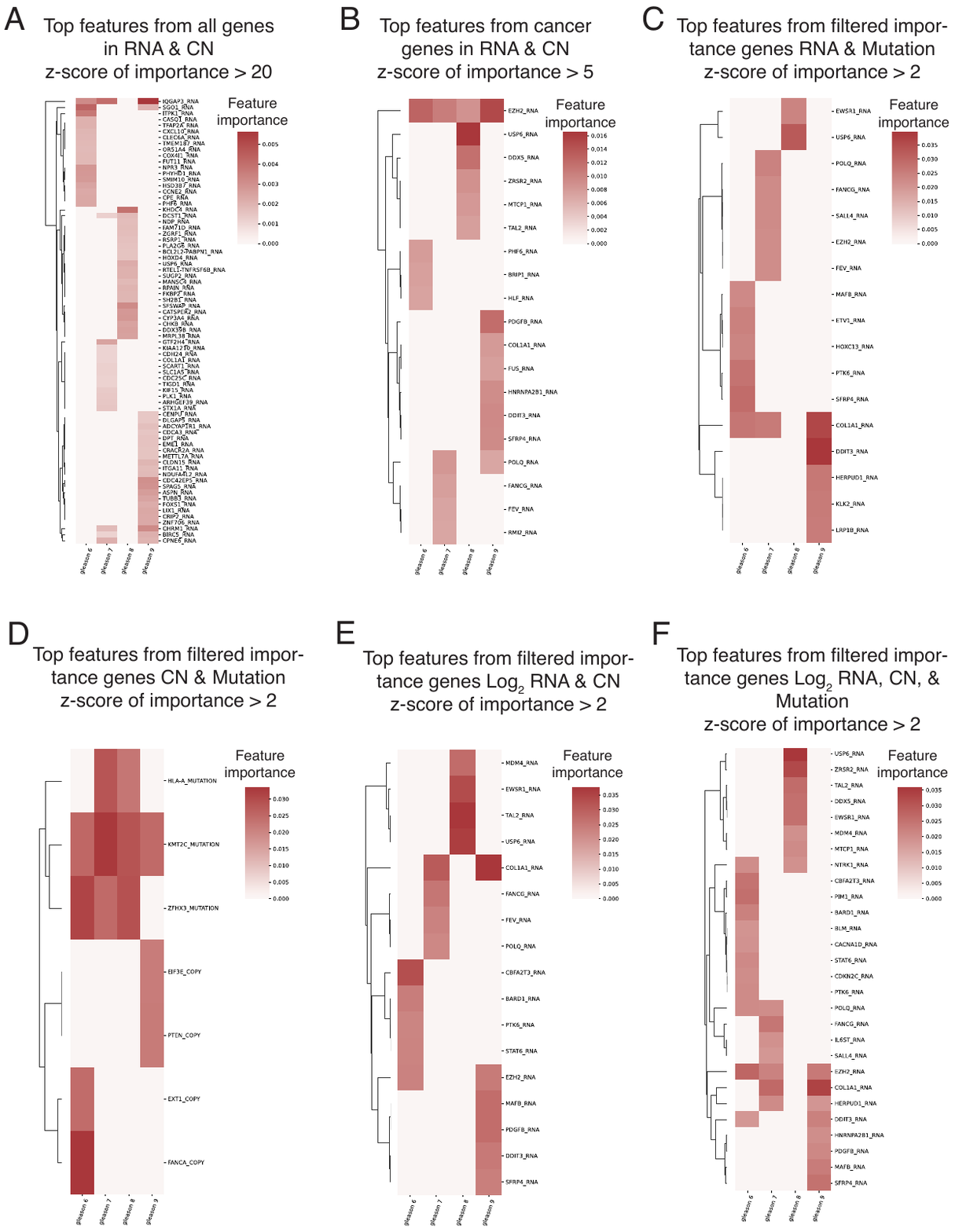}
\caption{Top features from a balanced binary classification using random forest. Cells are colored according to their feature importance score. Top features from RF analysis of A) all genes $log_2$ RNA expression and copy number (CN), B) cancer genes $log_2$ RNA expression and CN, C) z-score filtered genes $log_2$ RNA expression and mutation data, D) z-score filtered genes CN and mutation data, E) z-score filtered genes $log_2$ RNA expression and CN, F) z-score filtered genes $log_2$ RNA expression and CN and mutation data.}
\label{fig:imp_multi_heatmaps}
\end{figure}

\begin{table}[ht]
\caption{Important features from the random forest using all genes and RNA-seq data. Genes were found with a z-score threshold of 10.} 

\centering 
\resizebox{0.6\textwidth}{!}{\begin{tabular}{|c|c|c|} 
\hline\hline 
Gene & Important Gleason Scores & Importance Score (in order) \\ 
\hline 
BGN & 6 & 0.0075\\ [0.5ex] 
\hline
CENPU & 9 & 0.0029\\ [0.5ex]
\hline
CENPA & 6, 7, 9 & 0.0040, 0.0034, 0.0022\\ [0.5ex]
\hline
TACC3 & 6,7,9 & 0.0062, 0.0030, 0.0039\\ [0.5ex]
\hline 
PEBP4 & 7, 9 & 0.0015, 0.0052\\ [0.5ex]
\hline
ASF1B & 7 & 0.0013\\ [0.5ex]
\hline
MMP26 & 6 & 0.0022\\ [0.5ex]
\hline
CDK1 & 9 & 0.0017\\ [0.5ex]
\hline
ACP3 & 9 & 0.0019\\ [0.5ex]
\hline
\end{tabular}}
\label{table:rf_allgenes_imp_tab} 
\end{table}

\subsection{Biomarkers identified by Random Forest}

Given transcriptomic data and the entire gene feature space,  the random forest classifier model found hundreds of genes that were important for Gleason scores 6-9. Looking at genes that had a z-score > 10, genes such as \textit{BGN}, \textit{CENPU}, \textit{TACC3}, \textit{PEBP4}, \textit{ASF1B}, \textit{MMP26}, \textit{CDK1}, and \textit{ACP3} were among the most significant by importance scores (Table \ref{table:rf_allgenes_imp_tab}).
We also observed that in multi-omics experiments the importance of transcriptomic features were much greater than that of mutational load or CN. 
Across the multi-omics experiments, \textit{EZH2}, \textit{COL1A1}, \textit{USP6}, \textit{SFRP4}, \textit{DDIT3}, \textit{EZH2}, \textit{EWSR1}, \textit{TAL2}, \textit{KMT2C}, and \textit{ZFHX3} had the highest performances across Gleason scores (Figure \ref{fig:imp_multi_heatmaps}, Table \ref{table:bio_tab}). Of these genes, \textit{COL1A1} expression consistently had the highest importance for predicting Gleason 7 and 9, and  \textit{USP6} had the highest importance for predicting Gleason 8. \textit{FANCA} copy number and \textit{ZFHX3} mutational load both had significant importance for predicting Gleason 6.

\subsection{Enrichment Analysis}
To identify whether there is an association between prostate cancer Gleason scores and particular biological processes, we used two methods for set enrichment analysis (SEA)~\cite{da_wei_huang_bioinformatics_2009} to discover enriched gene sets amongst the most important genes found by the random forest. SEA shifts the focus from individual genes to relevant gene groups, allowing for greater identification of  biological processes affecting the phenotype. First, we applied a hypergeometric test to capture gene sets that are over-represented among the various features. Second, we utilized Gene Set Enrichment Analysis (GSEA) to analyze differential expression of gene sets from the transcriptomic data to distinguish between Gleason scores~\cite{subramanian_gene_2005}. Shared gene sets found from hypergeometric test and GSEA analysis would validate hypergeometric test results on transcriptomic data.

\subsection{Hypergeometric test results}

We considered first the Hallmark gene sets from the MSigDB for over-represented gene sets from important genes selected by the random forest classifier in \textit{all gene} experiments~\cite{liberzon_molecular_2011}. Only two gene sets consistently had significant $p$-values(<0.05) across all z-score thresholds: \textit{G2M Checkpoint} and \textit{E2F targets}. Specifically, these gene sets only appeared as significant for experiments using transcriptomic data. G2M checkpoint had its lowest $p$-value of 0.002 for Gleason 7 in the experiments using importance z-score thresholds of 5 and 20. \textit{G2M checkpoint} was also over-represented in genes significant for Gleason score 6 and 9 across multiple z-score thresholds. \textit{E2F targets} had a significant $p$-value of 0.040 for Gleason 7 in experiments using genes above z-score thresholds of 5 and 15. WNT Beta-Catenin pathway had a significant $p$-value of 0.050 for Gleason 9 in mutation data experiments with a z-score threshold of 15. 

\begin{figure}[!t]
\centering
\includegraphics[width=7.5in]{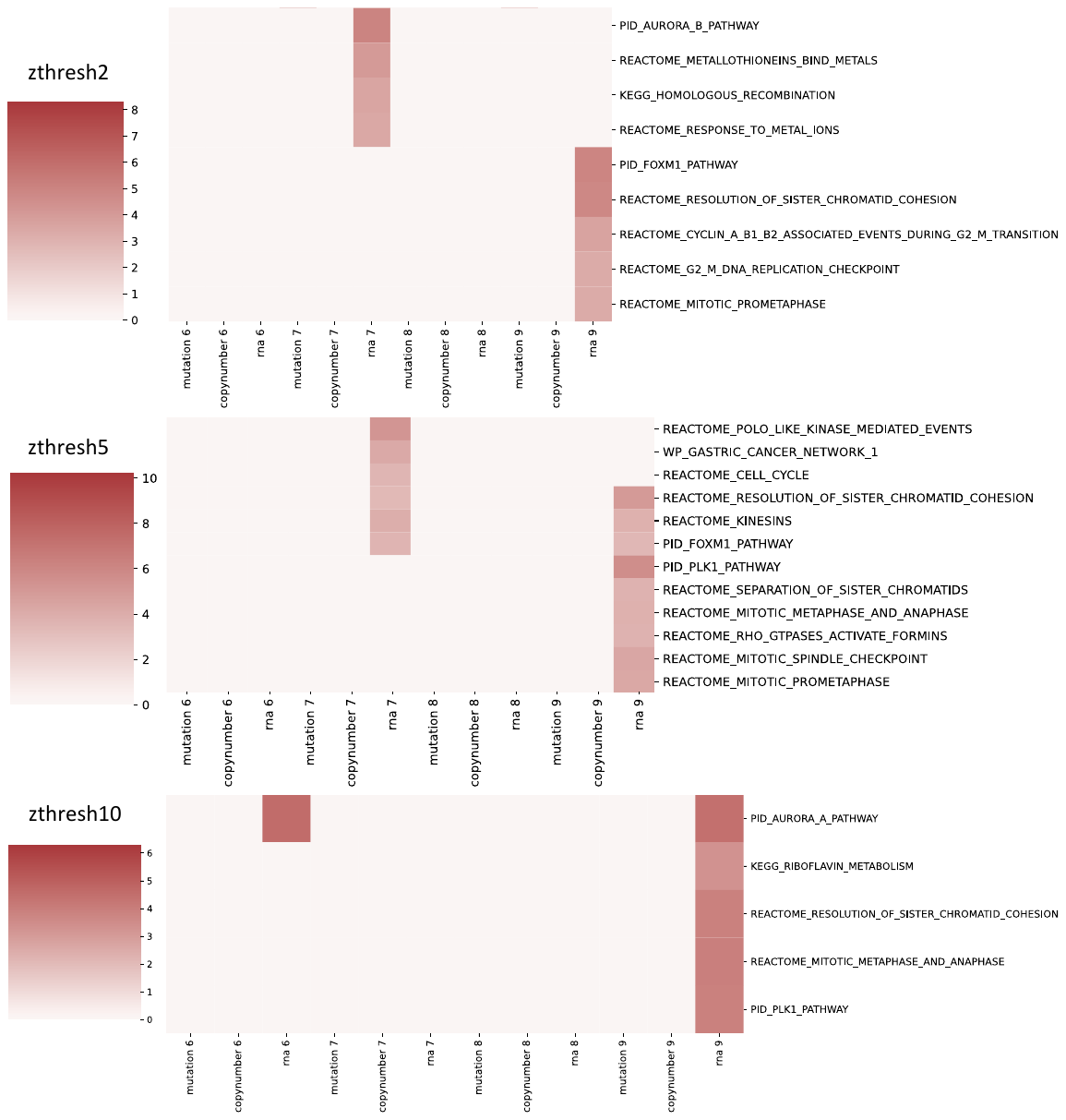}
\caption{ \textit{Z}-scores of top pathways from hypergeometric test experiments with C2CP gene sets. Genes whose feature importance z-score were higher than indicated z-score thresholds (2, 5, 10) were used in separate hypergeometric tests experiments. Columns indicate which experimental data and Gleason score were analyzed.  Only pathways that were significant in at least one comparison are shown. Cells are colored according to -log10($p$-value).}
\label{fig:c2cp_all_gene_hypergeo}
\end{figure}

When performing the hypergeometric test on important genes from \textit{all gene} experiments using the C2CP gene sets, we identified many pathways related to the cell cycle that had $p$-values significant for Gleason score 9, including the \textit{PID FOXM1 Pathway}, \textit{Reactome Resolution of Sister Chromatid Cohesion}, \textit{Reactome Mitotic Metaphase and Anaphase}, \textit{Reactome Mitotic Prometaphase}, and \textit{PID PLK1 Pathway} gene sets (Figure \ref{fig:c2cp_all_gene_hypergeo}). These sets were significant across feature importance z-score thresholds of 2, 5, and 10, showing how their signal persists as more significant genes are used for the hypergeometric function (Figure \ref{fig:c2cp_all_gene_hypergeo}).

\begin{table}[ht]
\captionsetup{justification=centering}
\caption{Important Hallmark gene sets from GSEA with FDR $q$-value less than 0.15.} 

\centering 
\resizebox{0.85\textwidth}{!}{\begin{tabular}{|c|c|c|c|} 
\hline\hline 
Gene set & Significant Gleason scores & NES scores (in order) & FDR \textit{q}-value 
(in order) \\ 
\hline 
\textit{G2M Checkpoint} & 6v9, 7v8, 7v9 & 1.9768, 1.8497, 2.0778 & 0.0259, 0.0588, 0.0119\\[0.5ex] 
\hline
\textit{E2F targets} & 6v8, 6v9, 7v8, 7v9 & 1.8574, 2.0400, 1.8699, 2.0706 & 0.0682, 0.0301, 0.0909, 0.0060\\ [0.5ex]
\hline
\textit{Myogenesis} & 6v7, 7v8 & -1.3967, -1.8477 & 0.1301, 0.1472\\ [0.5ex]
\hline
\textit{Spermatogenesis} & 6v8 & 1.8704 & 0.0868\\ [0.5ex]
\hline 

\end{tabular}}
\label{table:gsea_HM_comparison_imp} 
\end{table}

\begin{table}[ht]
\caption{Important C2CP gene sets from GSEA comparison experiment of Gleason scores 7 versus 9 with FDR \textit{q}-value less than 0.15 and NOM \textit{p}=value equal to 0.
}

\centering 
\resizebox{0.7\textwidth}{!}{\begin{tabular}{|c|c|c|c|} 

\hline\hline 
Gene set & NES & FDR \textit{q}-val \\ 
\hline 
PID FOXM1 Pathway & 2.1116  & 0.0942\\ 
\hline
WP Gastric Cancer Network 1 & 2.0914  & 0.0847\\ [0.5ex]
\hline
Reactome Resolution of Sister Chromatid Cohesion &2.0752  & 0.0788\\ [0.5ex]
\hline
KEGG Cell Cycle & 2.0645  & 0.073\\ [0.5ex]
\hline 
Reactome Cyclin A B1 B2 Associated Events During G2M Transition & 2.0585  & 0.0656\\ [0.5ex]
\hline
Reactome HDR Through Homologous Recombination HRR & 2.0504  & 0.0648\\ [0.5ex]
\hline
Reactome Resolution of D Loop Structures & 2.0488  & 0.0583\\ [0.5ex]
\hline
PID PLK1 Pathway & 2.0466  & 0.0531\\ [0.5ex]
\hline
WP Cell Cycle & 2.0349  & 0.0555\\ [0.5ex]
\hline
Reactome Cell Cycle Checkpoints & 2.0244  & 0.0594\\ [0.5ex]
\hline
Reactome Mitotic G1 Phase and G1 S Transition & 2.0219  & 0.056\\ [0.5ex]
\hline
Reactome Cyclin D Associated Events in G1 & 2.0165  & 0.0555\\ [0.5ex]
\hline
Reactome Fanconi Anemia Pathway& 2.0118  & 0.0554\\ [0.5ex]
\hline
Reactome Kinesins & 1.9862  & 0.0738\\ [0.5ex]
\hline
WP Retinoblastoma Gene in Cancer & 1.9813  & 0.0735\\ [0.5ex]
\hline
WP DNA Replication & 1.9771  & 0.0718\\ [0.5ex]
\hline
WP G1 to S Cell Cycle Control & 1.9757  & 0.0688\\ [0.5ex]
\hline
Reactome G1 S Specific Transcription & 1.9694  & 0.0707\\ [0.5ex]
\hline
Reactome G0 and Early G1 & 1.9682  & 0.068\\ [0.5ex]
\hline
KEGG DNA Replication & 1.9252  & 0.0753\\ [0.5ex]
\hline
Reactome G2M Checkpoints & 1.9173  & 0.0749\\ [0.5ex]
\hline
Reactome Separation of Sister Chromatids & 1.8955  & 0.0807\\ [0.5ex]
\hline
Reactome Mitotic Prometaphase & 1.8917  & 0.0818\\ [0.5ex]
\hline
PID E2F Pathway & 1.8485  & 0.0987\\ [0.5ex]
\hline
Reactome G2 M DNA Damage Checkpoint & 1.8455  & 0.0999\\ [0.5ex]
\hline
\end{tabular}}
\label{table:gsea_c2cp_binary_comp_imp_tab} 
\end{table}

\subsection{GSEA Experiments}

We ran pairwise comparisons of Gleason scores using GSEA to identify which biological processes may be differentially regulated between Gleason scores. Using Hallmark gene sets and selecting for those with significant FDR (False Discovery Rate) $q$-values (<0.15), we found that cell cycle related pathways \textit{Hallmark G2M checkpoint} and \textit{Hallmark E2F targets} were consistently up-regulated in a higher grade Gleason score when compared to a lower grade (Table \ref{table:gsea_HM_comparison_imp}). We further found that \textit{Hallmark Spermatogenesis} was up-regulated in Gleason 6 versus Gleason 8 cohorts, and \textit{Hallmark Myogenesis} was down-regulated in higher grade Gleason scores. The only C2CP gene sets to reach FDR $q$-values < 0.15 were cell cycle pathways up-regulated for Gleason 9 in Gleason 7 versus Gleason 9. These include cell cycle pathways such as \textit{PID PLK1 Pathway}, \textit{Reactome Resolution of Sister Chromatid Cohesion}, \textit{PID FOXM1 Pathway}, and \textit{Reactome Mitotic Prometaphase} (Table \ref{table:gsea_c2cp_binary_comp_imp_tab}). With both approaches, we find that cell cycle pathways are up-regulated in higher grade prostate cancers.

To identify differentially expressed pathways that may be specific to a particular Gleason score, we performed one-vs-all GSEA experiments. This approach discovers similar significant Hallmark gene sets (FDR $q$-value < 0.15) found in the pairwise comparisons, including \textit{Hallmark G2M checkpoint}, \textit{Hallmark E2F targets}, and \textit{Hallmark Myogenesis}. \textit{Hallmark G2M checkpoint} and \textit{Hallmark E2F targets} were down-regulated in Gleason 6 and Gleason 7 while up-regulated in Gleason 9 (Table \ref{table:gsea_1vsall_c2cp+hm_imp_tab}). \textit{Hallmark Myogenesis} was down-regulated in Gleason 8 compared to other Gleason scores. With C2CP gene sets, there was overlap between significant gene sets identified in GSEA one-vs-all comparisons and analysis from both the GSEA pairwise comparisons and the hypergeometric tests. These pathways included \textit{PID PLK1 Pathway}, \textit{Reactome Resolution of Sister Chromatid Cohesion}, \textit{Reactome Mitotic Metaphase and Anaphase}, \textit{Reactome Mitotic Prometaphase}, \textit{PID FOXM1 Pathway}, \textit{Reactome Mitotic Prometaphase}, \textit{G2M DNA Damage Checkpoint}, and \textit{E2F Pathway}. Though these significant pathways only appeared when testing Gleason 7 and Gleason 9, they followed the same trend reviewed in previous experiments where there was up-regulation for Gleason 9 and down-regulation for Gleason 7. 

\begin{figure}[!t]
\centering
\includegraphics[width=4in]{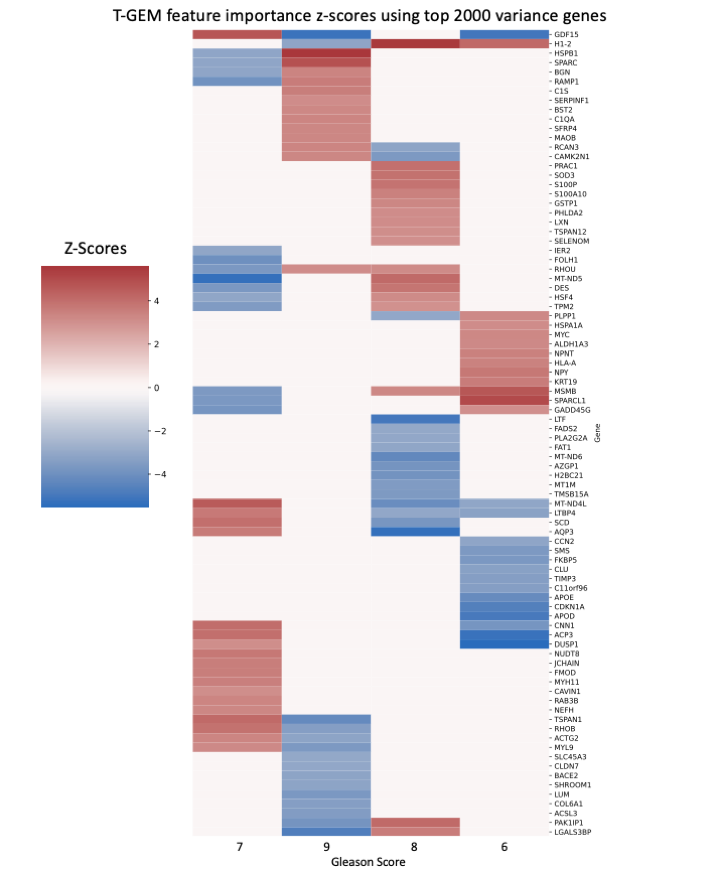}
\caption{Top features from T-GEM. Using the top 2000 genes by expression variance as input, features with an absolute z-score $\ge 2$ are plotted according to their z-score.  Positive or negative expression are expressed in red and blue, respectively. }
\label{fig:tgem_imp_heatmap}
\end{figure}

\begin{figure}[!t]
\centering
\includegraphics[width=5in]{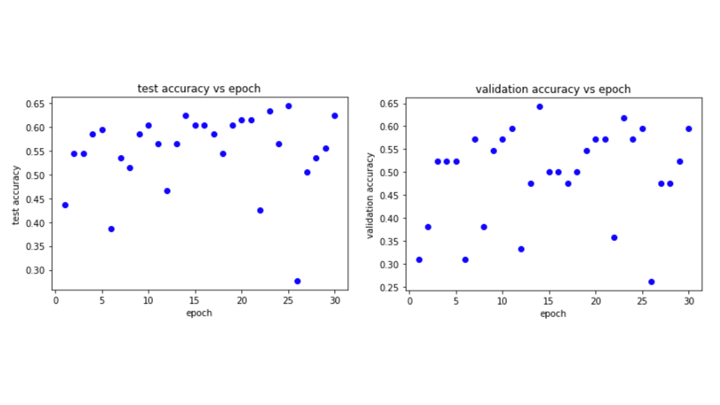}
\caption{Test and validation accuracy of T-GEM model over epochs 1 to 30.}
\label{fig:tgem_performance_plot}
\end{figure}
\subsection{T-GEM results}

To further expand the set of biomarkers associated to specific Gleason scores, we utilized an AI model based on transformers. Advanced deep learning models can have challenges in accurately fitting to the unique characteristics of gene expression data because biological data is unordered and contains complicated gene-gene relationships. Moreover, the "black box" nature of these models limits the interpretability of their results \cite{zhang_transformer_2022}. We applied an interpretable deep learning model for modeling gene expression data, Transformer for Gene Expression Modeling (T-GEM), which was developed for predicting cancer types. The T-GEM model has been shown to achieve an accuracy of 94.92\%, a Matthews correlation coefficient of 0.9469, and an AUC of 0.9987, outperforming models such as Random Forest, SVM, and CNN (Autokeras), in predicting cancer types \cite{zhang_transformer_2022}. Furthermore, the T-GEM model is able to discover gene pathways specific to cancer phenotypes and provide attention to specific cancer-related genes.

In this study, we modified T-GEM to predict Gleason scores rather than cancer type and provided the top 2000 genes by expression variance. During testing, we found that T-GEM performance was highly variable and did not effectively converge during training. The model achieved the highest performance during epoch 25, reaching a test accuracy of 0.644 and a validation accuracy of 0.595 (Figure \ref{fig:tgem_performance_plot}). The T-GEM model identified genes such as \textit{BGN}, \textit{SPARC}, \textit{RAMP1}, \textit{C1QA}, \textit{MAOB}, \textit{SERPINF1}, \textit{RHOU}, \textit{CAMK2N1}, \textit{HSPB1}, \textit{C1S}, \textit{BST2}, \textit{RCAN3}, and \textit{SFRP4} as positive markers for Gleason 9 (Figure \ref{fig:tgem_imp_heatmap}). Other genes, such as \textit{GDF15}, \textit{H1-2}, \textit{AQP3}, \textit{TSPAN1}, and \textit{ACP3} were found to have high positive and negative importances across Gleason 6-9.

 \section{Discussion}

\begin{table}
    \caption{Significant gene sets from GSEA binary experiments with FDR $q$-value less than 0.15. All gene sets that do not have Hallmark in name are gene sets from the C2CP collection.} 
    \centering
    \resizebox{0.9\textwidth}{!}
    {\begin{tabular}{|c|c|c|c|}
    \hline
         \textbf{Gene set} & \textbf{Gleason scores} & \textbf{NES scores (in order)} & \textbf{FDR\textit{q}-val (in order)}\\
\hline
\hline
         PID FOXM1 Pathway & 7, 9 & -2.0366, 2.1334 & 0.1301, 0.1306\\
         \hline
         Reactome Resolution of Sister Chromatid Cohesion & 7, 9 & -2.0337, 2.0326 & 0.1023, 0.0746\\
         \hline
         Reactome Mitotic Spindle Checkpoint& 7, 9 & -1.9554, 1.9490 & 0.1125, 0.0890\\
         \hline
         Reactome Mitotic Prometaphase & 7, 9 & -1.9492, 1.9130 & 0.1012, 0.0813\\
         \hline
         PID PLK1 Pathway & 7, 9 & -1.9123, 2.0114 & 0.1129, 0.0822\\
         \hline
         Reactome G2M Checkpoints & 7, 9 & -1.8239, 1.8561 & 0.1331, 0.1126\\
         \hline
         Reactome Mitotic Metaphase and Anaphase & 7, 9 & -1.7696, 1.8611 & 0.1439, 0.1170\\
         \hline
         PID E2F Pathway & 7, 9 & -1.7639, 1.8837 & 0.1435, 0.0944\\
         \hline
         Hallmark G2M Checkpoint & 6, 7, 9 & -2.0297, -1.9288, 2.0019 & 0.0228, 0.0239, 0.0240\\
         \hline
         Hallmark E2F Targets & 6, 7, 9 & -1.9694, -1.9388, 2.0308 & 0.0285, 0.0433, 0.0370\\
         \hline
         Hallmark Myogenesis & 8 & -1.9352 & 0.0460\\
         \hline
    \end{tabular}}
    \label{table:gsea_1vsall_c2cp+hm_imp_tab}
\end{table}

\begin{table}[ht]
\captionsetup{justification=centering}
\caption{Significant potential biomarkers from cancer gene experiments} 

\centering 
\resizebox{0.7\textwidth}{!}{\begin{tabular}{|c|c|c|c|c|} 
\hline\hline 
Gene & Gleason & Most informative & Best Experiment \\ 
 & Score(s) & data type & (z-score, Gleason score) \\
\hline 
\textit{COL1A1} & 7, 9 & RNA & LogRNA (12.0, 7) \\ [0.5ex] 
\hline
\textit{DDIT3} & 9 & RNA & RNA (7.3, 9) \\ [0.5ex]
\hline
\textit{EWSR1} & 8 & RNA & RNA + Mutation + CN + z-score filtering (3.0, 8)\\ [0.5ex]
\hline
\textit{EZH2} & 6,7,9 & RNA & RNA + CN (11.5, 9)\\ [0.5ex]
\hline 
\textit{FANCA} & 6 & CN & LogCN (6.1,6)\\ [0.5ex]
\hline
\textit{FANCG} & 6, 7 & RNA & RNA + mutation + CN + z-score filtering (3.1, 7) \\ [0.5ex]
\hline
\textit{KMT2C} & 6,7,8,9 & Mutation & Mutation + CN + z-score filtering (5.0, 7) \\ [0.5ex]
\hline
\textit{POLQ} & 6,7,9 & RNA & RNA + CN (7.2, 7) \\ [0.5ex]
\hline
\textit{SALL4} & 7 & RNA & RNA + Mutation (2.1, 7)\\ [0.5ex]
\hline
\textit{SFRP4} & 6,9 & RNA & RNA + CN (6.9, 9) \\ [0.5ex]
\hline
\textit{TAL2} & 8 & RNA & RNA + CN (4.1, 8)\\ [0.5ex]
\hline
\textit{TP53} & 7,8,9 & Mutation & Mutation (7.1, 9) \\ [0.5ex]
\hline
\textit{USP6} & 8 & RNA & RNA + mutation + CN + z-score filtering (4.7, 8) \\ [0.5ex]
\hline
\textit{ZFHX3} & 6,7,8 & Mutation & Mutation (9.8, 8) \\ [0.5ex] 
\hline

\end{tabular}}
\label{table:bio_tab} 
\end{table}

Our aim was to develop interpretable machine learning models to predict prostate cancer severity using Gleason scores and to discover associated biomarkers from omics data, whether using single or multiple data modalities. We found that using RNA-seq data, our models were able to predict Gleason 7 and 9 well, while mutational load data was able to predict Gleason 6. Copy number data was consistently comparable to the top-performing data type for each Gleason score. Depending on the data type and Gleason score being predicted, filtering the input data to the set of cancer genes was able to improve performance. Moreover, model performance for all Gleason scores was improved when combining data types. We observed that many of the important genes identified by the RF model have been previously found as known biomarkers for prostate cancer severity, and our RF model also discovered additional potential biomarkers.

\subsection{Significant biomarkers across single- and multi-omic random forest experiments}

Throughout the random forest experiments,  \textit{COL1A1},  \textit{FANCA},  \textit{FANCG},  \textit{ZFHX3},  \textit{USP6},  \textit{SALL4},  \textit{POLQ},  \textit{KMT2C},  \textit{EZH2},  \textit{SFRP4}, and  \textit{TP53} had some of the highest importance values for predicting Gleason scores, and these genes have all been identified as potential biomarkers of prostate cancer prognosis and severity \cite{giri_prevalence_2022, nientiedt_high_2020, sun_additive_2015, sun_frequent_2005, lange_genome-wide_2009, jedinak_et_al._abstract_2017, hayajneh_collagen_2018, liu_bisphenol_2022, he_inhibition_2016, kuei_dna_2020, rao_hijacked_2015, limberger_kmt2c_2022, henrich_usp6_2018, canesin_cytokines_2020, he_wnt3a:_2015, bernreuther_secreted_2020, sandsmark_sfrp4_2017, teroerde_revisiting_2021, xu_ezh2_2012, duan_ezh2:_2020}. Within these genes,  \textit{COL1A1},  \textit{ZFHX3}, and  \textit{USP6} were particularly significant. 
\textit{ZFHX3} was distinctly associated with Gleason 6 in single-omics experiments using mutational load data and continued to have importance for Gleason 6 in multi-omics experiments (Figures \ref{fig:imp_heatmaps}, \ref{fig:imp_multi_heatmaps}).  \textit{ZFHX3}, which is also known as \textit{ATBF1}, has been shown to be a tumor suppressor in prostate cancer in multiple studies and losses of it is a strong sign of uncontrolled prostate cancer growth \cite{sun_additive_2015, sun_frequent_2005, lange_genome-wide_2009}.  \textit{ZFHX3} has frequent deletions and mutations in prostate cancer, which the random forest model confirmed as \textit{ZFHX3}'s mutational load was more informative compared to its expression and copy number\cite{sun_additive_2015}. 

\textit{COL1A1} was among the strongest biomarkers the random forest model found, consistently performing at the top when predicting Gleason 7 and Gleason 9 across both single- and multi-omics experiments(Figures \ref{fig:imp_heatmaps}, \ref{fig:imp_multi_heatmaps}).  \textit{COL1A1} (collagen type I alpha 1 chain) has been shown to be a potential biomarker of prostate cancer in multiple studies, which have found elevated  \textit{COL1A1} expression levels in prostate cancer compared to Benign Prostatic Hyperplasia, exceptionally high \textit{COL1A1} expression in biochemical recurrence of prostate cancer, significant \textit{COL1A1} up-regulation in BPS-treated PC-3 cells, and definitive oncogenic properties \cite{jedinak_et_al._abstract_2017, hayajneh_collagen_2018, liu_bisphenol_2022, li_col1a1:_2022}. While there remains uncertainty as to the exact role of \textit{COL1A1} in prostate cancer, we have validated its use as a prostate cancer biomarker, and specifically may serve as a biomarker for Gleason 7 and 9.

We observed Gleason 8 was poorly predicted when using a single data type. However, a multi-omics approach significantly boosted the performance. \textit{USP6} was consistently a top performing gene for predicting Gleason 8 in multiple experiments that combined RNA data with other data types using z-score filtered cancer genes (Figure \ref{fig:imp_multi_heatmaps}).  \textit{USP6} has been found to promote tumorigenesis through the JAK1-STAT3 and Wnt/$\beta$-catenin pathways \cite{henrich_usp6_2018}. Persistent activation of JAK/STAT signaling correlates with tumor growth and disease progression in prostate cancer \cite{canesin_cytokines_2020}. The Wnt/$\beta$-catenin pathway, the canonical \textit{Wnt} signaling pathway, is responsible for stimulating tumor progression in multiple cancers, including prostate cancer \cite{he_inhibition_2016, he_wnt3a:_2015}. Through the support of various literature, there is a strong case for \textit{COL1A1},  \textit{ZFHX3}, and  \textit{USP6}, the top performing genes found by the RF, as biomarkers for prostate cancer severity.

Our random forest model, which captured known prostate cancer biomarkers, identified additional potential biomarkers for different Gleason scores, including \textit{TAL2},  \textit{EWSR1}, and  \textit{DDIT3} that warrants future study. Studies suggest that 9q34 chromosome duplication may be linked to prostate cancer and  \textit{TAL2} is a candidate for a prostate cancer gene from the 9q chromosome \cite{lange_genome-wide_2009}. As \textit{TAL2} had significant feature importance only when the RF model was given copy number data combined with other data types, the multi-omic model may have potential to capture known copy number markers and identify these features within multiple omic modalities. \textit{EWSR1} has been found to make a protein that can cooperate with the ERG transcription factor protein to promote prostate cancer \cite{10.1093/narcan/zcab033}. 
The gene \textit{SPOP} triggers  \textit{DDIT3} degradation, and mutations of  \textit{SPOP} that are linked with prostate cancer are defective in  \textit{DDIT3} degradation \cite{zhang_destruction_2014}. \textit{TAL2}, \textit{EWSR1}, and \textit{DDIT3} have not had their relation with prostate cancer prognosis and severity thoroughly investigated. However, based on their high feature importance in the results of this study, further research should be conducted in the future to confirm how they affect prostate cancer severity.

While we show there is little impact on performance when filtering using cancer genes, we tested the space of all genes to discover additional genes relevant to prostate cancer. Considering genes using RNA data, the random forest model identified several that are up-regulated in prostate cancer, including  \textit{CENPU},  \textit{CENPA},  \textit{TACC3},  \textit{PEBP4},  \textit{ASF1B},  \textit{MMP26},  \textit{CDK1}, and  \textit{ACP3}. These genes were not a part of the common cancer gene list from the CGC, yet have literature supporting their overexpression within prostate cancer, highlighting the potential for machine learning models to detect additional relevant prostate cancer genes from the complete ~20,000 gene space \cite{jean_m._winter_mapping_2016, qie_tacc3_2020, luo_roles_2019, han_knockdown_2018, cheng_mmp26:_2017, orlic-milacic_resolution_2013, saha_role_2020, han_prostate_2021, noauthor_cdk1_nodate}.

\subsection{Set enrichment analysis}

We performed set enrichment experiments to identify the greater biological pathways and mechanisms that could be primary drivers behind the prostate cancer phenotype and potentially capture biological processes over-represented within the important genes the RF found.

Across set enrichment experiments utilizing either MSigDB's Hallmark or C2CP gene sets, cell cycle gene sets were consistently over-represented. \textit{G2M checkpoint} and \textit{E2F targets} gene sets were significantly over-represented among random forest features across multiple z-score thresholds for Gleason 6, 7, and 9. In binary GSEA experiments, \textit{G2M checkpoint} and \textit{E2F targets} were down-regulated for Gleason 6 and 7, yet up-regulated for Gleason 9 (\ref{table:gsea_1vsall_c2cp+hm_imp_tab}). Furthermore, similar behaviors were observed in these differentially-expressed pathways from pairwise GSEA comparisons (Table \ref{table:gsea_HM_comparison_imp}). These patterns suggest both \textit{G2M checkpoint} and \textit{E2F targets} are positively correlated with increased severity of prostate cancer (Table \ref{table:gsea_1vsall_c2cp+hm_imp_tab}). 

\textit{G2M checkpoint} up-regulation in higher severity prostate cancer aligns with the pathway's biological function. The \textit{G2M checkpoint} pathway prevents cells from entering mitosis when DNA is damaged, providing an opportunity for repair and stopping the proliferation of damaged cells \cite{stark_analyzing_2004}. \textit{G2M checkpoint} pathway genes would have higher expression in higher grade prostate cancer, where there is an increased need to mitigate the growth of cancer cells. In studies of breast cancer, higher \textit{G2M checkpoint} pathway activity was correlated with enriched tumor expression of other cell proliferation-related gene sets, highlighting the enrichment of \textit{G2M checkpoint} in more aggressive cancers \cite{oshi_g2m_2020}.

Similarly, \textit{E2F} transcription factors regulate the cell cycle through the activation of genes important for the G1 to S phase cell cycle transition and are also involved in the activation of cell cycle regulation, DNA replication, DNA repair, DNA damage and G2/M checkpoints, chromosome transactions, and mitotic regulation \cite{ren_e2f_2002}. 
\textit{E2F} transcription factors have been shown to significantly affect the aggressiveness of prostate cancer, where increased \textit{E2F} gene expression had a strong association with greater risk of death \cite{foster_transcription_2004}. Furthermore, \textit{E2F targets}-related genes, including \textit{PLK1}, have high prognosis value for prostate cancer and high-risk groups formed from an \textit{E2F targets} gene signature demonstrate poor disease outcome, resistance to treatments, immunosuppression, and abnormal growth characteristics \cite{xia_novel_2022}.

We observed enrichment of additional cell cycle gene sets from the C2CP database in the top features from the random forest, including PID FOXM1 Pathway, Reactome Resolution of Sister Chromatid Cohesion, Reactome Mitotic Metaphase and Anaphase, Reactome Mitotic Prometaphase, and PID PLK1 Pathway, for Gleason 9 in hypergeometric tests and GSEA experiments (Figure \ref{fig:c2cp_all_gene_hypergeo}, Tables \ref{table:gsea_c2cp_binary_comp_imp_tab} and \ref{table:gsea_1vsall_c2cp+hm_imp_tab}). 
The \textit{FOXM1} transcription factor has been found to promote tumorigenesis by promoting cell cycle progression through direct proliferation-driving targets like c-Myc (MYC) \cite{katzenellenbogen_targeting_2023}. Furthermore, the \textit{FOXM1} transcription factor has been found to be highly expressed in prostate cancer cells and has been shown to promote prostate cancer progression by regulating PSA gene transcription \cite{liu_foxm1_2017}.
Overexpression of \textit{PLK1} has been found to override mitotic checkpoints, which can lead to immature cell division with aneuploidy, and also contributes to cancer development by promoting excessive cell proliferation through the dysregulation of checkpoint functions \cite{lee_polo-like_2014}. \textit{PLK1} regulates proper spindle assembly and chromosome segregation, while the inhibition of \textit{PLK1} has been shown to lead to greater effectiveness of cancer treatment\cite{mao_plk1_2018, shin_cotargeting_2019, kalous_multiple_2023}.

The over-representation of cell cycle gene sets within top features of the random forest from both Hallmark and C2CP gene sets highlights the importance of these pathways in prostate cancer progression. 
These enrichment results are further supported by GSEA experiments. GSEA shared the same gene space as hypergeometric test, which encompassed the entire prostate cancer dataset without any cancer gene-related filters. After analysis on all 19,000+ genes, the aforementioned cell cycle gene sets were shown to be significantly differentially expressed. Specifically, these pathways tended to be over-expressed in high grade prostate cancer. The agreement between the hypergeometric test and GSEA highlights the cell cycle pathways' value as biomarkers of prostate cancer severity and may represent potential targets for therapeutic development.

\subsection{Biomarkers from transformer based analysis} 

T-GEM was leveraged as an alternative model for discovering prostate cancer biomarkers from expression data. It was able to identify many genes positively associated to Gleason 9, including  \textit{BGN},  \textit{SPARC},  \textit{RAMP1},  \textit{C1QA},  \textit{MAOB},  \textit{SERPINF1},  \textit{RHOU},  \textit{CAMK2N1},  \textit{HSPB1},  \textit{C1S},  \textit{BST2},  \textit{RCAN3}, and  \textit{SFRP4} (Figure \ref{fig:tgem_imp_heatmap}). 6 of 13 genes (\textit{BGN},  \textit{SPARC},  \textit{MAOB},  \textit{RHOU},  \textit{HSPB1}, and  \textit{SFRP4}) have been shown to be overexpressed in high severity prostate cancer, and \textit{RAMP1} has been shown to have high expression in prostate cancer overall \cite{bernreuther_secreted_2020, sandsmark_sfrp4_2017, jacobsen_up-regulation_2017, lopez-moncada_sparc_2022, tainjie_pu_abstract_2022, vasiljevic_association_2013, de_piano_lipogenic_2020, logan_ramp1_2013}. \textit{SFRP4} was also identified by the random forest to be a potential biomarker for Gleason 9, further supporting its role as a marker for higher prostate cancer severity (Figures \ref{fig:imp_heatmaps} and\ref{fig:imp_multi_heatmaps}). 

T-GEM further identified genes associated to the other Gleason scores, including  \textit{AQP3}, \textit{TSPAN1}, \textit{GDF15}, \textit{MYC}, and \textit{ACP3} (Figure \ref{fig:tgem_imp_heatmap}). \textit{AQP3} has been found to increase prostate cancer cell motility and invasion\cite{chen_aquaporin_2015}. \textit{TSPAN1} has been shown to be driven by androgen in prostate cancer and increases cell survival and motility, which would lead to the spread of the cancer \cite{munkley_cancer-associated_2017}. \textit{GDF15} plays a critical role in the development of prostate cancer bone metastasis \cite{siddiqui_gdf15_2022}.  \textit{MYC} is a known oncogene and contributes to the development of prostate cancer \cite{qiu_myc_2022}.   \textit{ACP3} encodes prostatic acid phosphatase (PAP), which is a marker that can be used to diagnose and monitor prostate cancer\cite{han_prostate_2021}. Furthermore, \textit{ACP3} mRNA levels could be used to identify prostate cancer subtypes \cite{han_prostate_2021}.  \textit{ACP3} was also identified by the all genes random forest classifier to be an important biomarker for Gleason 9 (Table \ref{table:rf_allgenes_imp_tab}). 

We do recognize that T-GEM experiments were inconsistent, with test and validation accuracy widely varying between epochs. As neural network models typically require a great deal of training data, the 500 prostate cancer samples included within the TCGA-PRAD dataset may not have been enough to sufficiently train the T-GEM model. A future line of work would be to study whether the use of Generative Adversarial Networks (GAN) designed to generate more training samples could be beneficial. Through the use of GANs, the T-GEM model would have greater available training cases and could possibly improve its performance.

Despite T-GEM's performance challenges that were likely the result of the  limited sample size, this neural network approach was still able to identify biomarkers for Gleason scores with supporting evidence from literature for their association to prostate cancer. It was also able to recover biomarkers also discovered by the random forest, including  \textit{ACP3},  \textit{MYC}, and  \textit{SFRP4} (Figures \ref{fig:imp_heatmaps} and \ref{fig:tgem_imp_heatmap}, Table \ref{table:rf_allgenes_imp_tab}). These genes should be further studied as potential therapeutic treatments in prostate cancer.  Our results demonstrate the promise of neural network approaches to find biomarkers provided there is sufficient data.

\section{Conclusion}

Our study aimed to discover potential biomarkers for predicting prostate cancer Gleason scores, using two different machine learning approaches along with clinical and multi-omic data. 
We found that individual data types were able to predict particular Gleason scores successfully and validated several top ranking biomarkers in the literature. Moreover, by combining datasets together, we were able to identify biomarkers that went unnoticed when using a single data type. By combining different approaches and analyses we found multiple genes, such as \textit{COL1A1} and \textit{SFRP4}, and cell cycle pathways, such as \textit{G2M checkpoint},  \textit{E2F targets}, and the PLK1 pathway, that were important predictive features for particular Gleason scores. The combination of these approaches shows the potential for easier, unbiased grading of prostate cancers, and for greater understanding of the biological processes behind prostate cancer severity that could provide novel therapeutic targets.

\section*{Acknowledgments}

I would like to thank my science teacher, Ms. Jennifer Doran, at Rye Country Day School for her time and for help.

\bibliography{citations}

\bibliographystyle{plainurl}

\end{document}